\documentclass[journal]{IEEEtran}
\usepackage{latexsym,,amssymb,amsmath,graphicx,epsf,cite,bbm,float}
\usepackage{ifpdf}
\usepackage{epstopdf,mathtools}

\usepackage{algorithm,algorithmic}
\usepackage{amsmath,amssymb,bm}
\usepackage{amsfonts,dsfont,color,bbm,subcaption}

\def\ninept{\def\baselinestretch{1}}
\ninept

\newcommand{\abs}[1]{|#1|}

\DeclareMathOperator*{\argmax}{arg\,max}

\DeclareMathOperator*{\median}{median}

\usepackage{hyperref,amsthm}

\newtheorem{remark}[]{Remark}

\begin{document}

\title{Blind Source Separation for Mixture of Sinusoids with Near-Linear Computational Complexity} 
\author{\IEEEauthorblockN{Kaan Gokcesu}, \IEEEauthorblockN{Hakan Gokcesu} }
\maketitle

\begin{abstract}
	We propose a multi-tone decomposition algorithm that can find the frequencies, amplitudes and phases of the fundamental sinusoids in a noisy observation sequence. Under independent identically distributed Gaussian noise, our method utilizes a maximum likelihood approach to estimate the relevant tone parameters from the contaminated observations. 
	When estimating $M$ number of sinusoidal sources, our algorithm successively estimates their frequencies and jointly optimizes their amplitudes and phases. Our method can also be implemented as a blind source separator in the absence of the information about $M$. The computational complexity of our algorithm is near-linear, i.e., $\tilde{O}(N)$.
\end{abstract}

\section{Introduction}

In many fields of signal processing such as
voice analysis \cite{fischer1954acoustic,titze1994toward,drioli2003emotions}, power system analysis \cite{grainger1999power,lovisolo2005efficient,akers2006hydraulic}, vibration analysis \cite{norton1990fundamentals,petyt2010introduction,brandt2011noise}, communication systems \cite{stremler1990introduction,schwartz1995communication,tse2005fundamentals}, biomedical signal processing \cite{gokcesu2018adaptive,akay2012biomedical,gokcesu2018semg,chang2010biomedical}, anomaly detection \cite{chandola2009anomaly,gokcesu2017online,patcha2007overview,delibalta2016online,ahmed2016survey,gokcesu2018sequential}; the accurate estimations of frequencies, amplitudes and phases for a  multi-tone signal corrupted by random noise is a fundamental problem \cite{luo2015frequency}. 
The parameter estimations of a mixture of sinusoids is a prominent topic of research since it arises in many real world applications such as wireless communications \cite{luo2015frequency,luo2016interpolated}, radar imaging \cite{selva2005efficient,wang2016parameters} and spectroscopy analysis \cite{umesh1996estimation,duda2011dft}. Due to noise, robust approaches \cite{steinhardt2018robust,gokcesu2021generalized,gokcesu2022nonconvex} are needed.

In time series analysis \cite{gokcesu2021nonparametric,hamilton2020time}, the estimation of multi-tone sinusoidals has been studied extensively, which has given rise to many various algorithmic approaches \cite{stoica1997introduction,zielinski2011frequency}, These algorithms can be categorized into two distinct types, which are the parametric and nonparametric approaches.

Nonparametric techniques include many traditional methods such as
Capon Method \cite{capon1969high}, Amplitude and Phase EStimation (APES) \cite{li1996adaptive} and Iterative Adaptive Approach (IAA) \cite{yardibi2010source}, which can be utilized to estimate the frequencies of the sinusoidal components in the signal by picking the peaks of the spectral estimates provided by the algorithms without knowing the number of tones. The spectra estimations of these methods are potentially highly accurate. They can estimate the multi-tone components in $O(N^2+N'\log N')$ time complexity, where $N$ is the number of signal samples and $N'$ is the length of the estimated spectra \cite{glentis2008fast,xue2011iaa,angelopoulos2012computationally}. However, for accurate detection by picking the peaks, $N'$ needs to be substantially higher than $N$, which makes this method significantly costly and quadratic at best, i.e., $\Omega(N^2)$.

Parametric estimation techniques \cite{gokcesu2017density} are also utilized as an alternative to the nonparametric spectra estimation. The application of this approach results in the aim of finding estimates for the sinusoidal parameters. The parametric approaches in the literature can be divided into two main categories, which are the time domain and frequency domain approaches \cite{belega2010accuracy}.

The time domain approaches utilize various subspace methods including Matrix Pencil (MP) \cite{hua1990matrix} and Estimation of Signal Parameters via Rotational Invariant Techniques (ESPRIT) \cite{vanhuffel1994algorithm,haardt1995unitary}. The most famous adaptation is MUltiple SIgnal Classification (MUSIC) algorithm \cite{schmidt1986multiple}, which is a generalization of the Pisarenko's Method \cite{pisarenko1973retrieval}. These approaches heavily utilize Singular Value Decomposition (SVD) \cite{wall2003singular} to separate the observed signal space into true signal subspace and noise subspace. Another approach is the iterative optimization techniques, which includes Iterative Quadratic Maximum Likelihood (IQML) \cite{bresler1986exact} and Weighted Least Squares (WLS) \cite{chan2010efficient,sun2012efficient}, both of which minimize the error between estimated and observed signals iteratively under various constraints. 

Nonetheless, the time domain methods, much like the nonparametric ones, suffer from high computational complexity, which results from the costly matrix operations such as inversion, Eigen-Decomposition (ED) \cite{abdi2007eigen} and SVD. These operations generally have a complexity of $O(N^3)$ and they are $\Omega(N^2)$ \cite{boyd1994linear}. This issue results from the fact that the estimation of the fundamental frequencies is relatively hard, i.e., the optimization of the frequencies is a form of non-convex optimization problem \cite{henrion2004solving,gokcesu2021regret,gokcesu2022low}. To address the high costs, the work in \cite{selva2005efficient} proposed an efficient approach based on maximum likelihood by utilizing the Vandermonde structure \cite{klinger1967vandermonde}. However, relatively high computational cost may still be present depending on the number of tones in the signal.

To this end, frequency domain approaches are much more popular because of their efficient computation, which is possible due to methods such as Fast Fourier Transform (FFT) \cite{nussbaumer1981fast}. These include the CLEAN \cite{gough1994fast} and RELAXation maximum likelihood algorithm (RELAX) \cite{li1996adaptive} methods. They utilize an iterative procedure of estimation (by maximizing the periodogram) and subtraction. However, the direct maximization over periodogram may result in the order of $O(N^{-1})$ frequency estimation error \cite{aboutanios2011estimating}. To increase accuracy, denser periodograms may be used (e.g., by zero-padding), which will multiplicatively increase the complexity by the reciprocal of the density parameter, i.e., $\epsilon^{-1}$. Many algorithms \cite{quinn1994estimating,aboutanios2005iterative,yang2011noniterative,selva2011efficient} have been designed to address this. However, since they are for the single tone estimation problem, their application to multi-tone settings may result in poor performance because of cross interferences. 
These interferences result from the Picket Fence Effect (PFE) and Spectral Leakage Effect (SLE) \cite{santamaria1998improved,zhang2001algorithm}.
Hence, there is a need to jointly optimize the multi-tone parameters \cite{ye2017rapid}.
The reduction of interference have been studied extensively \cite{diao2013interpolation,duda2014interpolated,duda2011dft,luo2015frequency,luo2016interpolated}, where the observation is analyzed after multiplying with a tapering window \cite{gokcesu2022smoothing,prabhu2014window}. However, the incorporation of non-rectangle windows may be detrimental to the frequency estimation accuracy \cite{ye2017rapid}.

\section{Preliminaries}\label{sec:prelim}
We have observed samples $\boldsymbol{x}=\{x_n\}_{n=0}^{N-1}$. Let these be noisy observations of the true signal samples $\boldsymbol{\bar{x}}=\{\bar{x}_n\}_{n=0}^{N-1}$. Hence,
\begin{align}
	x_n=\bar{x}_n+v_n,&&\forall n.
\end{align}
where $v_n$ is additive white Gaussian noise.
Let the noise-free samples $\boldsymbol{\bar{x}}$ be from a multi-tone signal, i.e., a mixture of sinusoids with varying amplitudes, phases and frequencies. Hence,
\begin{align}
	\bar{x}_n=\sum_{m=1}^{M}A_m\sin(w_mn+\theta_m),
\end{align}
for amplitudes $\{A_m\}_{m=1}^M$, phases $\{\theta_m\}_{m=1}^M$ and frequencies $\{w_m\}_{m=1}^M$ for some $M$. 

\begin{remark}
	The parameter $M$, i.e., the number of sinusoid in the signal can be known or unknown. We may also only have an access to an upper-bound $M_0$.
	
\end{remark}

First, let us look at a simpler version of the problem where we know $M$. If we want to estimate the sinusoids in a least squares sense, we have the following optimization problem:
\begin{align}
	\min_{\{\hat{A}_m,\hat{\theta}_m,\hat{w}_m\}_{m=1}^M}\sum_{n=0}^{N-1}\left(x_n-\sum_{k=1}^{K}\hat{A}_m\sin\left(\hat{w}_mn+\hat{\theta}_m\right)\right)^2.
\end{align}

Although the estimation of the amplitudes $\hat{A}_m$ and $\hat{\theta}_m$ given the frequencies $\hat{w}_m$ is relatively easier; the estimation of $\hat{w}_m$ makes the problem much harder. If we had known the frequencies $\hat{w}_m$, this would be a convex optimization problem; however, when we also want to estimate the frequencies, we encounter a non-convex problem setting.

To this end, we first tackle the problem of estimating the frequencies $w_k$. To detect these tone frequencies, we need to study the Discrete Time Fourier Transform (DTFT), since the least squares frequency solution is the peak of DTFT. The calculation of the whole transform is exhaustive. However, we can calculate it for any single frequency $w$ in $O(N)$ time, where we have
\begin{align}
	\boldsymbol{X}(w)=\sum_{n=0}^{N-1}x_ne^{-jwn}.
\end{align}
Hence, the question is where to sample it.
For this reason, we can start by taking a discrete fourier transform (DFT) $\boldsymbol{X}=\{X_k\}_{k=0}^{K}$, where $K=\lfloor N/2\rfloor$ and
\begin{align}
	X_k=\sum_{n=0}^{N-1}x_ne^{-j\frac{2\pi kn}{N}}.
\end{align}
Hence, in a sense, DFT is a uniformly sampled version of DTFT. The brute force approach to calculate DFT takes $O(N)$ time per each $X_k$, which is not better than DTFT. However, with efficient FFT methods, the whole sequence can be calculated in $O(N\log N)$ linearithmic time.

After the calculation of the DFT values, we will estimate the fundamental frequencies $\hat{w}_m$ and estimate their amplitude and phases in succession. The complete approach is given in the next section.

\section{The Iterative Algorithm}
\subsection{Accurate Detection of the Fundamental Frequency Bin}\label{sec:bin}

To detect the fundamental frequency, one may find the peaks of DFT. However, this approach has a strong assumption that the fundamental frequency is an integer multiple of $2\pi/N$. In general, this does not have to be and we need to explore the DTFT of the signal. Nonetheless, we can utilize DFT since it is a uniformly sampled version of DTFT. 

Our goal is to find the fundamental frequency, i.e., most dominating sinusoidal in the signal. If we had access to the whole infinite length signal, this would be a delta function located at its frequency. However, we only have access to a limited number of samples. This coincides with observing the infinite length signal after multiplication with a rectangle window. When we multiply the original signal with a rectangle window, the frequency domain will be convolved with its Fourier.  

When $N$ is large, this convolved kernel is approximately a sinc function (ignoring the common amplitude), i.e.,
\begin{align}
	S(w)=\frac{\sin(\frac{Nw}{2})}{\frac{Nw}{2}}.
\end{align}
This signal is symmetric around $0$ and $S(0)=1$, $S(2\pi m /N)=0, \forall m\in\mathbbm{Z}\setminus{0}$.
Since the original signal spectra will be convolved with this function, when the fundamental frequency is an integer multiple of $2\pi/N$, its contribution on the other DFT values is zero. However, if it is not, there will be spectral leakage. To estimate the fundamental frequency, we utilize this spectral leakage. We observe that the main lobe of this function is of two sample distance length in DFT domain. Hence, using two consecutive DFT values, we can approximate this sinc behavior, and estimate the magnitude and the frequency of the fundamental sinusoidal. 

Let two consecutive DFT values be $X_k$ and $X_{k+1}$ for some $k$. Resulting from the convolution with sinc, we have the following

\begin{align}
	\abs{X_{k+1}}=& A_k{\frac{\sin(\pi \delta_k)}{\pi \delta_k}},\\
	\abs{X_{k}}=& A_k{\frac{\sin(\pi (1-\delta_k))}{\pi (1-\delta_k)}},
\end{align} 
where $A_k$ is the magnitude estimate of the fundamental sinusoidal with the frequency estimate $2\pi(k+1-\delta_k)/N$ and $0\leq\delta_k\leq1$.
Since $\theta_k\triangleq\sin(\pi\delta_k)=\sin(\pi(1-\delta_k))$, we have
\begin{align}
	\frac{1-\delta_k}{\delta_k}=&\frac{\abs{X_{k+1}}}{\abs{X_{k}}},\\
	\delta_k=& \frac{\abs{X_{k}}}{\abs{X_k}+\abs{X_{k+1}}}.
\end{align}
Thus,
\begin{align}
	A_k=\frac{\pi}{\theta_k}\frac{ \abs{X_{k}}\abs{X_{k+1}}}{\abs{X_{k}}+\abs{X_{k+1}}}.
\end{align}

We calculate all such $A_k$ from $k=0$ to $k=K-1$. Whichever $A_k$ is maximum, we explore that bin. Hence, finding the fundamental frequency bin will take $O(N)$ time. 

\subsection{Robust Bin Exploration for Fundamental Frequency}\label{sec:exp}
After deciding on the exploration bin, i.e., $k,k+1$; we assume that the amplitude inside that bin approximate part of a sinc function, hence, it has a bell shape symmetric around its peak. To find the peak, we can utilize the following successive sampling method:
\begin{enumerate}
	\item Set $w_l=2\pi k /N$ and $w_r=2\pi(k+1)/N$. Set $w_m=(w_l+w_r)/2$
	\item If $\abs{\boldsymbol{X}(w_l)}=\abs{\boldsymbol{X}(w_r)}$, STOP and return $w_m$;\\
	else, continue\label{step:ite1}
	\item Evaluate $\boldsymbol{X}(w_m)$.
	\item If $\abs{\boldsymbol{X}(w_l)}<\abs{\boldsymbol{X}(w_r)}$, set $w_l\leftarrow w_m$;\\
	else set $w_r\leftarrow w_m$
	\item Return to Step \ref{step:ite1}
\end{enumerate}
In each iteration, this algorithm will get closer to the peak of the symmetric bell, i.e., the fundamental frequency. However, it may never terminate since the stopping criterion is an equality. To this end we can consider an $\epsilon$ closeness such as $$-\epsilon\leq\abs{\boldsymbol{X}(w_l)}-\abs{\boldsymbol{X}(w_r)}\leq\epsilon.$$ 
However, such a closeness parameter may be arbitrarily dependent on the signal power. A better measure can be $$1-\epsilon\leq \frac{\abs{\boldsymbol{X}(w_l)}}{\abs{\boldsymbol{X}(w_r)}}\leq 1+\epsilon.$$ 
However, it is hard to limit the computational complexity with such a criterion. Thus, a more natural criterion is
$$w_r-w_l\leq \frac{2\pi}{N}\epsilon. $$
Hence, the algorithm takes the form:
\begin{enumerate}
	\item Set $w_l=2\pi k /N$ and $w_r=2\pi(k+1)/N$. Set $w_m=(w_l+w_r)/2$
	\item If $w_r-w_l\leq 2\pi\epsilon/N$, STOP and return $w_m$;\\
	else, continue\label{step:ite1.1}
	\item Evaluate $\boldsymbol{X}(w_m)$.
	\item If $\abs{\boldsymbol{X}(w_l)}<\abs{\boldsymbol{X}(w_r)}$, set $w_l\leftarrow w_m$;\\
	else set $w_r\leftarrow w_m$
	\item Return to Step \ref{step:ite1.1}
\end{enumerate}

\begin{remark}
	With this criterion, an $\epsilon$ close frequency can be reached in $O(\log(\epsilon^{-1}))$ evaluations.
\end{remark}

However, we observe that this algorithm heavily utilizes the symmetricity of the bell shape. Any outside interference such as leakage from other frequencies may invalidate it. Thus, we relax the condition on the bell shape such that instead of a strict symmetricity, we assume a strict quasi-concavity.

\begin{remark}
	When $\abs{\boldsymbol{X}(w)}$ is strictly quasi-concave over $w\in[w_l,w_r]$, we have the following properties:
	\begin{itemize}
		\item $\abs{\boldsymbol{X}(\lambda w_1+(1-\lambda)w_2)}>\min(\abs{\boldsymbol{X}(w_1)},\abs{\boldsymbol{X}(w_2)})$ for any $0<\lambda<1$ and for all $w_1,w_2\in[w_l,w_r]$.
		\item $\abs{\boldsymbol{X}(w)}$ has a unique maximizer $w^*$ in the set $[w_l,w_r]$. 
		\item For a set of frequencies $w_1<w_2<w_3$, if $\abs{\boldsymbol{X}(w_2)}>\max(\abs{\boldsymbol{X}(w_1)},\abs{\boldsymbol{X}(w_3)})$; the maximizer $w^*$ is in the set $(w_1,w_3)$.
		\item If $\abs{\boldsymbol{X}(w_1)}=\abs{\boldsymbol{X}(w_2)}$ for some $w_1<w_2$, we have $w^*\in(w_1,w_2)$.
	\end{itemize}
\end{remark}

To utilize quasi-concavity, we revise the algorithm as:
\begin{enumerate}
	\item Set $w_l=2\pi k /N$ and $w_r=2\pi(k+1)/N$. Set $w_m=(w_l+w_r)/2$. Evaluate $\boldsymbol{X}(w_m)$.
	\item If $\abs{w_l-w_r}\leq{2\pi}\epsilon/N$, STOP and return $w_m$;\\
	else, continue\label{step:ite2}
	\item Evaluate $\boldsymbol{X}(w_{lm})$ and $\boldsymbol{X}(w_{mr})$, where $w_{lm}=(w_l+w_m)/2$ and $w_{mr}=(w_m+w_r)/2$.
	\item Set $w_{max}=\argmax_{w\in\{w_l,w_{lm},w_m,w_{mr},w_r\}}(\abs{\boldsymbol{X}(w)})$
	\item If $w_{max}=w_m$, set $w_l\leftarrow w_{lm}$, $w_r\leftarrow w_{mr}$;\\
	else if $w_{max}\in\{w_l,w_{lm}\}$, set $w_m\leftarrow w_{lm}$, $w_r\leftarrow w_{m}$;
	else if $w_{max}\in\{w_{mr},w_{r}\}$, set $w_l\leftarrow w_{m}$, $w_m\leftarrow w_{mr}$
	\item Return to Step \ref{step:ite2}
\end{enumerate}

We observe that at each iteration of this algorithm, the separation between the frequencies $w_l$ and $w_r$ decrease by half. Hence, an $\epsilon$ close solution can be reach in $O(\log(\epsilon^{-1}))$ iterations and evaluations. 
Since we utilize the quasi concavity of the DTFT, we observe that it may be susceptible to noise. There may be evaluations that break the quasi-concavity property. To this end, for robustness, we need to preserve the quasi-concavity. We observe that the quasi-concavity criterion implies an ordering between the samples. To see this, let $w_1,w_2,w_3$ be three frequencies such that $w_1<w_2<w_3$ and $w_1,w_2,w_3\in\mathcal{W}$, there the DTFT is quasi-concave in $\mathcal{W}$. 
Because of quasi-concavity, we have $4$ distinct situations:
\begin{itemize}
	\item $\abs{\boldsymbol{X}(w_1)}\leq \abs{\boldsymbol{X}(w_2)}\leq \abs{\boldsymbol{X}(w_3)}$
	\item $\abs{\boldsymbol{X}(w_1)}\leq \abs{\boldsymbol{X}(w_3)}\leq \abs{\boldsymbol{X}(w_2)}$
	\item $\abs{\boldsymbol{X}(w_3)}\leq \abs{\boldsymbol{X}(w_1)}\leq \abs{\boldsymbol{X}(w_2)}$
	\item $\abs{\boldsymbol{X}(w_3)}\leq \abs{\boldsymbol{X}(w_2)}\leq \abs{\boldsymbol{X}(w_1)}$
\end{itemize}
These situations imply the following:
$$\abs{\boldsymbol{X}(w_2)}\geq \median(\abs{\boldsymbol{X}(w_1)}, \abs{\boldsymbol{X}(w_2)}, \abs{\boldsymbol{X}(w_3)}).$$
In the algorithm, whenever this implication is invalid, we can correct it by replacing the evaluated DTFT with the median of samples. Although this criterion needs to be satisfied for all evaluations, we can do it successively. Each corrected point will remain valid and we can just correct the newly sampled points. Thus, the algorithm takes the following final form:.

\begin{enumerate}
	\item Set $w_l=2\pi k /N$ and $w_r=2\pi(k+1)/N$. Set $w_m=(w_l+w_r)/2$. Evaluate $\boldsymbol{X}(w_m)$.
	\item If $\abs{\boldsymbol{X}(w_m)}< \median(\abs{\boldsymbol{X}(w_l)}, \abs{\boldsymbol{X}(w_m)}, \abs{\boldsymbol{X}(w_r)})$, set $\abs{\boldsymbol{X}(w_m)}\leftarrow \median(\abs{\boldsymbol{X}(w_l)}, \abs{\boldsymbol{X}(w_m)}, \abs{\boldsymbol{X}(w_r)})$
	\item If $\abs{w_l-w_r}\leq{2\pi}\epsilon/N$, STOP and return $w_m$;\\
	else, continue\label{step:ite3}
	\item Evaluate $\boldsymbol{X}(w_{lm})$ and $\boldsymbol{X}(w_{mr})$, where $w_{lm}=(w_l+w_m)/2$ and $w_{mr}=(w_m+w_r)/2$.
	\item If $\abs{\boldsymbol{X}(w_{lm})}< \median(\abs{\boldsymbol{X}(w_l)}, \abs{\boldsymbol{X}(w_{lm})}, \abs{\boldsymbol{X}(w_m)})$, set $\abs{\boldsymbol{X}(w_{lm})}\leftarrow \median(\abs{\boldsymbol{X}(w_l)}, \abs{\boldsymbol{X}(w_{lm})}, \abs{\boldsymbol{X}(w_m)})$
	\item If $\abs{\boldsymbol{X}(w_{mr})}< \median(\abs{\boldsymbol{X}(w_m)}, \abs{\boldsymbol{X}(w_{mr})}, \abs{\boldsymbol{X}(w_r)})$, set $\abs{\boldsymbol{X}(w_{mr})}\leftarrow \median(\abs{\boldsymbol{X}(w_m)}, \abs{\boldsymbol{X}(w_{mr})}, \abs{\boldsymbol{X}(w_r)})$
	\item Set $\boldsymbol{X}_{max}=\max_{w\in\{w_l,w_{lm},w_m,w_{mr},w_r\}}(\abs{\boldsymbol{X}(w)})$
	\item If $\boldsymbol{X}_{max}=\abs{\boldsymbol{X}(w_m)}$, set $w_l\leftarrow w_{lm}$, $w_r\leftarrow w_{mr}$;\\
	else if $\boldsymbol{X}_{max}\in\{\abs{\boldsymbol{X}(w_l)},\abs{\boldsymbol{X}(w_{lm})}\}$, set $w_m\leftarrow w_{lm}$, $w_r\leftarrow w_{m}$;\\
	else if $\boldsymbol{X}_{max}\in\{\abs{\boldsymbol{X}(w_{mr})},\abs{\boldsymbol{X}(w_{r})}\}$, set $w_l\leftarrow w_{m}$, $w_m\leftarrow w_{mr}$
	\item Return to Step \ref{step:ite3}
\end{enumerate}

Thus, we can explore the bin in $O(N\log(\epsilon^{-1}))$ time.

\subsection{Jointly Optimized Successive Tone Extraction}\label{sec:ext}
After estimating a fundamental frequency $w$, we extract it from the observed samples $\boldsymbol{x}$. We know that while our signal includes a sinusoidal of frequency $w$, we do not know its amplitude or phase. Fortunately, the estimation of the amplitude and phase is not nearly as hard as the estimation of the frequency. We can estimate them in a least squares manner. To do this, we first create two vectors $\boldsymbol{s_{w}}=\{s_{w,n}\}_{n=0}^{N-1}$ and $\boldsymbol{c_{w}}=\{c_{w,n}\}_{n=0}^{N-1}$, where
\begin{align}
	s_{w,n}=\sin(wn),&&c_{w,n}=\cos(wn), &&&\forall n.
\end{align}
Then, we model $\boldsymbol{x}$ as a linear combination of the vectors $\boldsymbol{s_w}$ and $\boldsymbol{c_{w}}$, i.e.,
\begin{align}
	\boldsymbol{\hat{x}}=\alpha\boldsymbol{s_w}+\beta\boldsymbol{c_{w}}.
\end{align}
Hence, we solve the following least squares problem
\begin{align}
	\min_{a,b\in\Re}\sum_{n=0}^{N-1}(x_n-\alpha s_{w,n}-\beta c_{w,n})^2.
\end{align}

The minimizers of this optimization problem has a closed form solution, which is given by
\begin{align}
	\begin{bmatrix}
		\alpha_*\\
		\beta_*
	\end{bmatrix}
	=
	\begin{bmatrix}
		\boldsymbol{s_w^Ts_w}& \boldsymbol{s_w^Tc_w}\\
		\boldsymbol{c_w^Ts_w}& \boldsymbol{c_w^Tc_w}
	\end{bmatrix}^{-1}
	\begin{bmatrix}
		\boldsymbol{s_w^Tx}\\
		\boldsymbol{c_w^Tx}
	\end{bmatrix}.
\end{align}
After acquiring $\alpha_*$ and $\beta_*$, we have
\begin{align}
	A_w=\sqrt{\alpha^2_*+\beta_*^2},&&\theta_w=\tan^{-1}\left(\frac{\beta_*}{\alpha_*}\right),
\end{align}
which gives the least square fit sinusoid $\boldsymbol{z_w}=\{z_{w,n}\}_{n=0}^{N-1}$ such that
\begin{align}
	z_{w,n}=A_w\sin(wn+\theta_w).
\end{align}
After acquiring $\boldsymbol{z_w}$, we extract it from the observation, i.e., $\boldsymbol{\tilde{x}}=\boldsymbol{x}-\boldsymbol{z_w}$. We do the same process to the residual. After finding the relevant fundamental frequency bin and its exploration, we have a new fundamental frequency $w'$. We do the same least squares fitting to the newly acquired frequency, however instead of just doing it for the residual we combine the new fundamental frequency with prior frequency estimations and jointly optimize them. Hence, if we have $m$ frequency estimations $\{w_{i}\}_{i=1}^m$, we create the relevant sinusoids $\boldsymbol{s_{w_i}}$ and $\boldsymbol{c_{w_i}}$ for all $i$. Then, we create the sinusoid matrix $\boldsymbol{Y_m}$ such that
\begin{align}
	\boldsymbol{Y_m}=
	\begin{bmatrix}
		\boldsymbol{s_{w_1}}&\boldsymbol{c_{w_1}}&\ldots& \boldsymbol{s_{w_m}}&\boldsymbol{c_{w_m}}.
	\end{bmatrix}
\end{align} 
Let the linear weights be $\boldsymbol{v}$ such that
\begin{align}
	\boldsymbol{\lambda_m^T}=
	\begin{bmatrix}
		\alpha_1&\beta_1&\ldots&\alpha_m&\beta_m
	\end{bmatrix}
\end{align}
Then, we have
\begin{align}
	\boldsymbol{\lambda_m}=(\boldsymbol{Y_m^TY_m})^{-1}\boldsymbol{Y_m^Tx}.
\end{align}
We create the residual $\boldsymbol{\tilde{x}}=\boldsymbol{x}-\boldsymbol{Y_m\lambda}$ and use it to find the next $(m+1)^{th}$ fundamental frequency. If we have $m$ current fundamental frequency estimations, the creation of the new correlation matrices will take additional $O(mN)$ time. Its inversion will take $O(m^3)$ time, which will make the total complexity $O(m^3+mN)$. If we do this for up to $M$ number of fundamental sinusoids, the complexity will be $O(M^2 N)$ assuming $M\leq O(\sqrt{N})$. 

\section{Conclusion}\label{sec:disc}
We proposed a multi-tone decomposition algorithm that can find the frequencies, amplitudes and phases of the fundamental sinusoids in a noisy observation sequence. Our approach adopts a maximum likelihood scheme under Gaussian noise and utilizes Fast Fourier Transform, robust quasi-concave maximization and least squares fitting to efficiently compute the estimations.
When estimating $M$ number of sinusoidal sources,
the FFT in the beginning of iterations will take $O(MN\log N)$ time; finding the fundamental frequency bin will take $O(MN)$ time; exploring the bin will take $O(MN\log(\epsilon^{-1}))$ time; the least squares fitting will take $O(M^2N)$ time. Thus, the total complexity is $O(MN(M+\log N+\log(\epsilon^{-1})))$. For blind separation without knowledge of $M$, if we set $\epsilon=O(1/N^p)$ for any finite $p$ and $M=O(\log^q N)$ for any finite $q$, the final complexity will be near-linear, i.e., $\tilde{O}(N)$.

\bibliographystyle{IEEEtran}
\bibliography{double_bib}
\end{document}